# A Quantum Router For The Entangled Web

*Bernardo A. Huberman\*, Bob Lund\*\**
*\*CableLabs, \*\*Occams Solutions*
\*[b.huberman@cablelabs.com](mailto:b.huberman@cablelabs.com)
\*\*[bob@occams.solutions](mailto:bob@occams.solutions)

### Abstract

Qubit transmission protocols are presently point-to-point, and thus restrictive in their functionality. A quantum router is necessary for the quantum Internet to become a reality. We present a quantum router design based on teleportation, as well as mechanisms for entangled pair management. The prototype was validated using a quantum simulator.

## INTRODUCTION

Recent advances in the generation and transmission of entangled photons - both in open space and through fiber optics - open the door to the deployment of a quantum network that will surpass the capabilities of the Internet in terms of security as well as access to quantum computation [KI]. The unit of information transfer will be the qubit, rather than the standard bit, and it will enable a number of novel quantum applications, some of which have been already developed. Among them, the secure distribution of cryptographic keys [QKD], a quantum advantage in communication complexity [QTMADV], the solution of a number of coordination problems among people and institutions in totally private fashion [HUB], the provision of public goods without free riding and central controls [PUB] and quantum auctions with privacy concerns that exceed the capabilities of classical ones [CHE].

Equally important, the quantum Internet will enable the access to future powerful quantum computers, which will be, of necessity, first available in the cloud.

Presently, the practical implementation of quantum teleportation allows for the transmission of qubits over very large distances using quantum repeaters [DUA]. Moreover, open space transmission from orbiting satellites has been demonstrated [PEN], which together with on-demand generation of solid-state quantum emitters [REI] opens the possibility of global quantum communication becoming a reality.

But the existing quantum network architecture is mostly based on point-to-point communication and optical switches that at best mimic the old patch networks of telephony. Just as those patch networks were eventually replaced by automatic routers in the Internet, we still need to create and deploy quantum routers that will enable the transmission of quantum teleported photons from any source to any destination.

This paper reports on the design and testing via quantum simulations of such a quantum router.

## THE ENTANGLED WEB

The fundamental enabler for the Entangled Web is the ability to transmit qubits over fiber optics [CWQUBIT]. Financial institutions are using this commercially for Quantum Key Distribution [QKD]. But transport of an instance of a quantum photon is limited to 100 to 200 miles, due to attenuation of the optical signals. The quantum "no-cloning" restriction [NOCLONE] prohibits the amplification of qubits as a means to extend the transmission distance. A second practical limitation of today's technology is the exchanges of photons over continuous wave (CW) fiber optics, which uses point-to-point links (Figure 1).

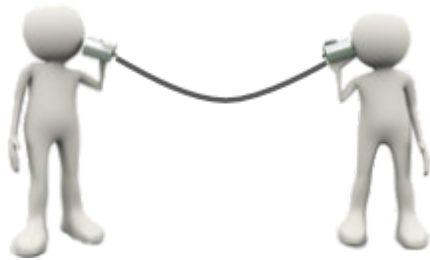

**Figure 1 – Point-to-point communication**

Quantum teleportation [QTP] has been shown to enable the transmission of qubits over a greater distance than is possible with CW optics [INDSRC] [TELEPORT]. Teleportation is a protocol whereby the quantum state of a particle, e.g. photon, is transferred from photon 1 to photon 2 via a second pair of entangled photons. Figure 2 shows the quantum teleportation protocol steps.

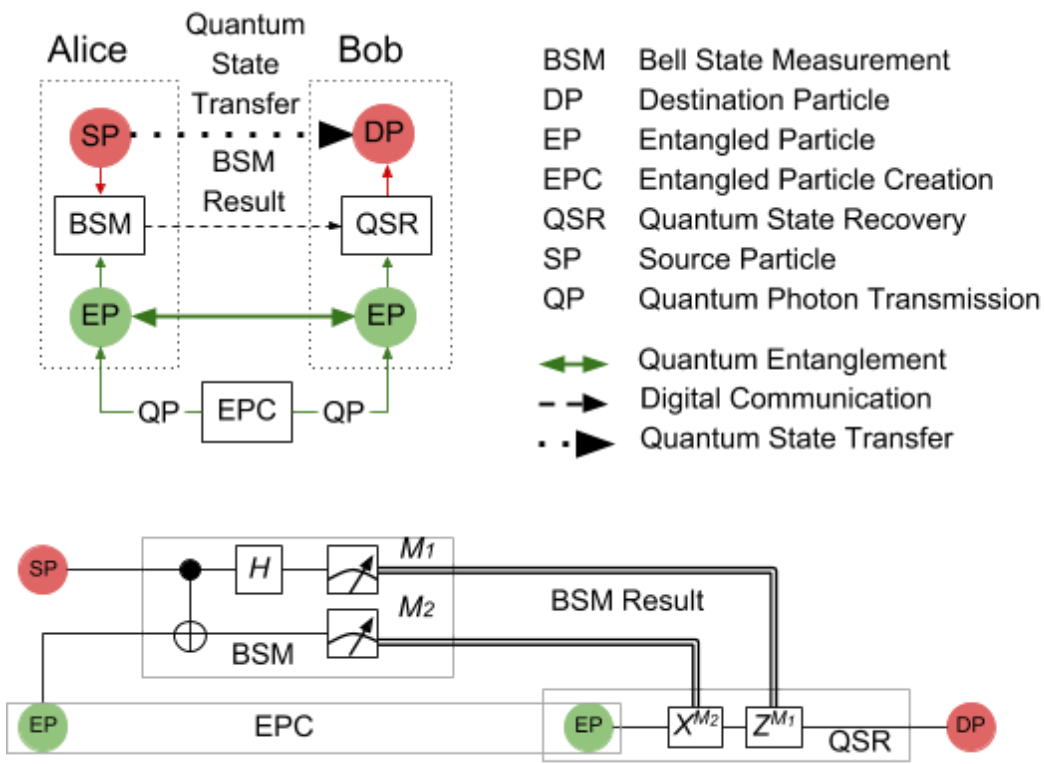

Source: Quantum Computation and Quantum Information p27

**Figure 2 - Quantum Teleportation**

1. A pair of entangled particles are created (EPC). One is sent to Alice and another to Bob over an continuous wave (CW) optical communication channel. This results in the creation of a quantum entangled channel between Alice and Bob.

2. Alice creates the Bell State with her source particle, whose state she wants to transfer to Bob, and her entangled particle. She then makes a Bell State Measurement (BSM) of her source and entangled particle. The measurement causes Alice's source and entangled particle to collapse into a basis state so they are no longer in superposition.

3. Alice sends the Bell State Measurement result to Bob over a digital communication channel.

4. Bob performs Quantum State Recovery operation, i.e. he applies quantum operations to his entangled particle, depending on the BSM result from Alice. The effect of this is a Quantum State Transfer from Alice to Bob. Bob now possesses a quantum particle with the same exact state as the one Alice used to possess.

The lower diagram in Figure 2 shows the quantum circuit for teleportation, and the distribution of the circuit across Alice, Bob and the entanglement creation.

The quantum teleportation protocol also enables the transfer of the state of an entangled quantum, known as "entanglement swapping" [ENTSWAP]. If Alice's source particle is entangled with some third party, e.g. Charlie, then teleportation will result in Charlie and Bob now possessing an entangled connection (Figure 3).

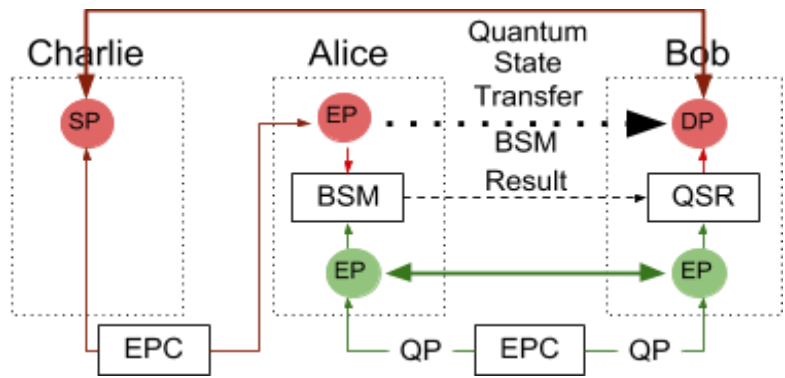

**Figure 3 – Entangled State Teleportation**

The transfer of entanglement enables cloud or distributed quantum computing, which are likely to be useful to quantum information processing due to the high cost and limited availability of early quantum computers.

Positioning the Entangled Particles Creation function midway between Alice and Bob allows for up to twice the transmission distance to be achieved, compared to a single CW optical link. Instances of the protocol can be repeated, in principle allowing transmission of quantum state over unlimited distances (Figure 4).

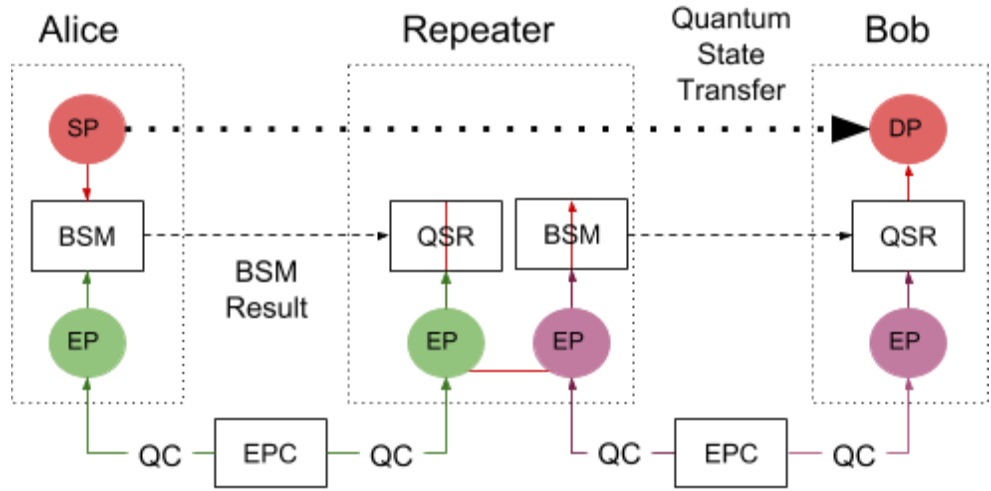

**Figure 4 – Quantum Repeater**

However, the system depicted in Figure 4, while better than Figure 1, still requires manual configuration, which won't scale for the Entangled Web (Figure 5).

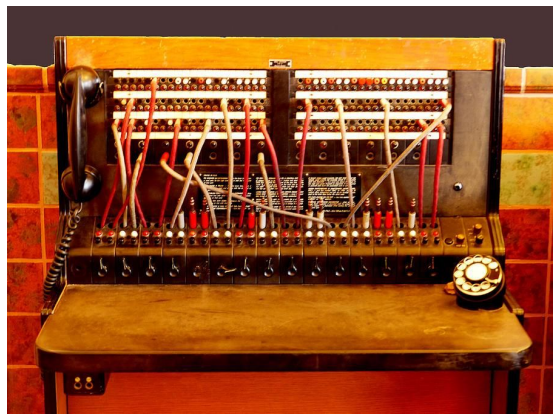

**Figure 5 – Limits of manual re-configuration**

As was the case with telephone and data networks, an automated control plane is required for the Entangled Web, i.e. a Quantum Router.

# A QUANTUM ROUTER FOR THE ENTANGLED WEB

## *Quantum Routing Architecture*

While quantum Repeaters can be used to transfer state across a number of hops between a source and destination, the route between source and destination is a static one. The Entangled Web requires a routing layer similar in functionality to today's digital Internet. Figure 6 illustrates the need for the additional routing capability so that a Source node can transfer qubits to a selected destination.

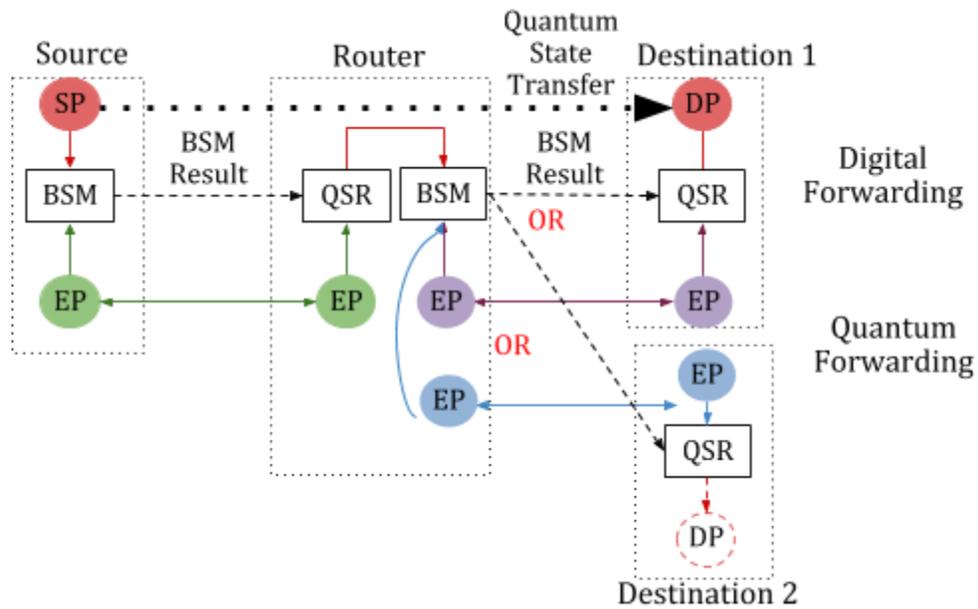

**Figure 6 - Routing in the Entangled Web**

A Quantum Router (QR) connects to multiple destinations via:

1. A classical communication path for sending Bell State Measurement Results (Digital Forwarding).
2. A quantum communication path via entanglement (Quantum Forwarding).

The QR makes a forwarding decision as to which classical and quantum paths to use, based on addressing information provided by the source, e.g. Destination 1 or Destination 2.

Figure 7 shows a more generalized example of the Entangled Web.

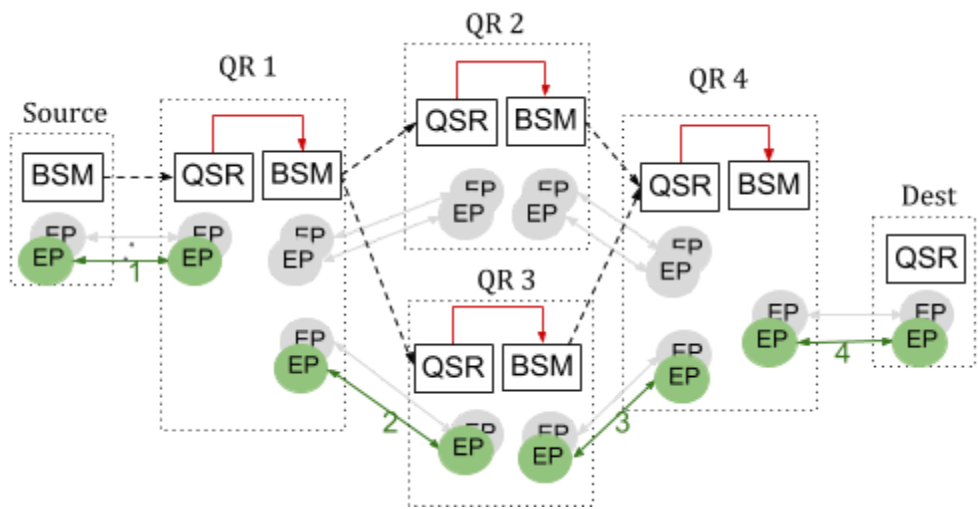

**Figure 7**

Forwarding the state of a quantum particle from Source to Dest involves applying the following steps at each node of the network:

1. Choose a forwarding interface (digital for the Bell State Measurement and entangled pair (for the quantum state) to the next hop.
2. Perform the Bell State Measurement with the qubit to be forwarded and the entangled pair to the next hop.
3. Forward the Bell State Measurement results to the next hop.

The green entangled pairs (EP) and associated links in Figure 7 show an example forwarding path for a quantum state from the Source to Dest. Each node in the network maintains a routing table to determine the next QR to which the quantum state is transfered. Table 1 shows the set of forwarding information for node QR1 in Figure 7. Note that the forwarding table is the same for digital and quantum forwarding even though the forwarding mechanism is different.

| Destination | Forwarding Interface | Link Metric |
|---|---|---|
| "Dest" | "QR2" | 10 |
| "Dest" | "QR3" | 1 |

**Table 1 – QR1 Forwarding Table**

## Quantum Router Design

Digital forwarding is what it is used in digital networks today, e.g. the IP network. Quantum forwarding via teleportation is completely new. The coordinated operation of Digital and Quantum forwarding requires information transfer between the digital and quantum forwarding planes. Figure 8 shows a system design that defines a simple interface between the quantum and digital planes so that existing digital forwarding can be leveraged for quantum forwarding. All quantum operations take place in the quantum network. Digital information necessary for quantum operations, e.g. which qubit to operate on and athe Bell State Measurement result, are communicated via a small set of commands between the digital and quantum networks. Forwarding information is communicated via messages in the digital network.

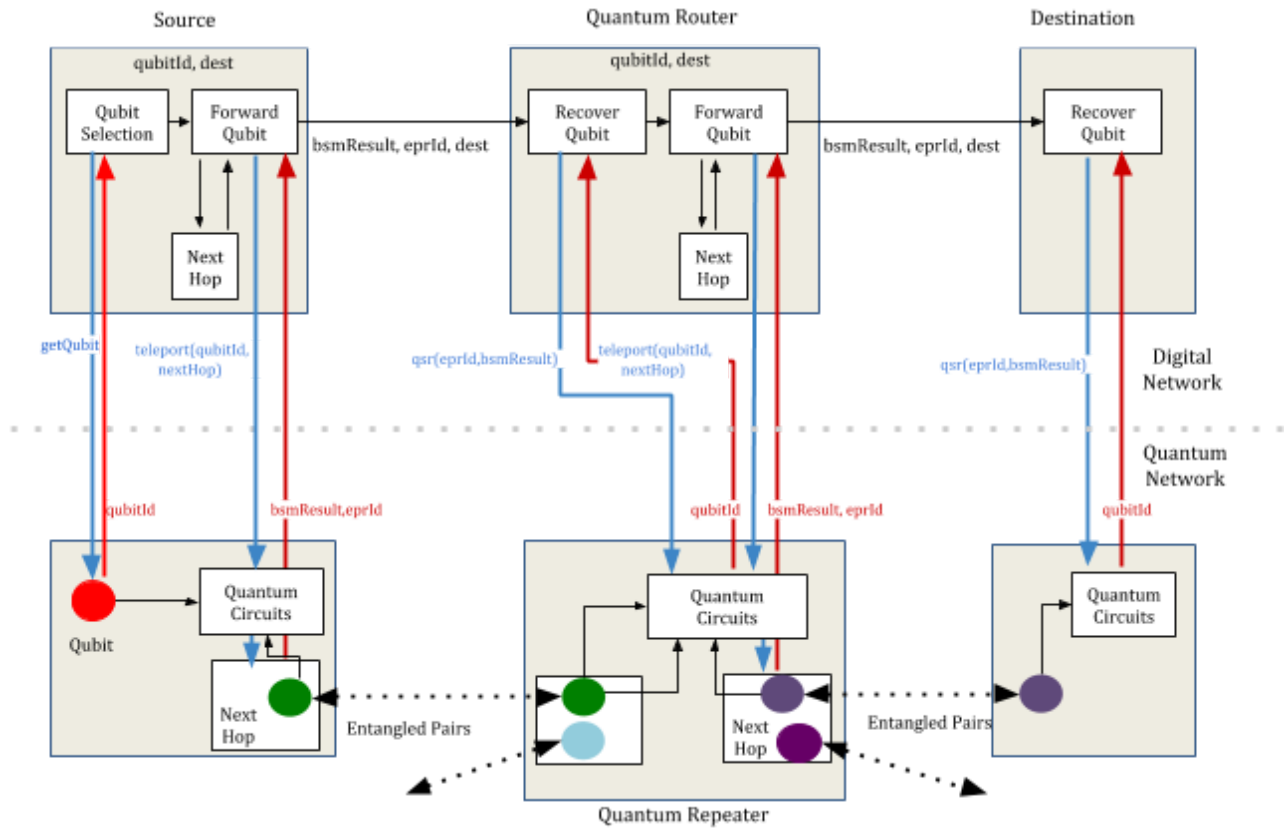

**Figure 8**

Figure 8 shows a simple network where a qubit state in the Source node is to be sent to the Destination node.

1.  A qubit selection process sends a *getQubits* request with a reference to the qubit to be transferred. This reference may create a new qubit with a specified state or reference an existing qubit. In either case, a reference to a qubit, *qubitId,* is returned.

2.  The *qubitId* and the Destination node identifier are sent to the Forward Qubit process. Forward qubit uses the Destination node identifier to determine the next hop, *nextHop.*

3.  The *teleport(qubitId, nextHop)* request is sent to the quantum network component associated with the Source node.
    a.  The Quantum Network uses the *nextHop* to get the *eprId* (entangled pair ID) of an entangled pair between the Source node and the Quantum Network component associated with the *nextHop* node. Management of entangled pairs is discussed in the Entangled Pair Management section.
    b.  A Bell State is created with the qubits referenced by *qubitId* and the qubit in *eprId* held by this node. A Bell State Measurement is made on these qubits. The *teleport* results (*eprId* and *bsmResult)* are returned to the Qubit Forwarding process in the Digital Network.

4.  The Forward Qubit process sends a forwarding message, *forward(Destination node identifier, eprId, bsmResult),* to the *nextHop* node.

5.  The Recover Qubit process in the receiving node in the Digital Network uses the forwarding message contents to recover the quantum state "forwarded" via teleportation by sending qsr*(eprId, bsmResult)* to its Quantum

Network component.

6. The quantum state recovery (qsr) portion of the teleportation protocol is executed, according to the *bsmResult,* on the qubit of the entangled pair referenced by *eprId*, which must reference the same entangled pair in the sending and receiving nodes. This is discussed further in the Entangled Pair Management section.

   At this point, the *eprId* qubit held in the receiving node contains the state of the qubit from the sending node. A *qubitId* referencing this qubit is returned to the Digital Network portion of the receiving node.

7. If the receiving node is a Quantum Router, the same operations as in the Source node are performed to forward the quantum state on to the Destination node.

   If the receiving node is the Destination, further operations on the qubit referenced by *qubitId* can be performed.

## Entangled Pair Management

An Entangled Pair Management (EPM) function manages the creation of the entangled pairs that comprise the quantum forwarding portion of the Entangled Web. The EPM does the following:

1. Creates entangled pairs.
2. Creates unique labels for the entangled pairs, i.e. the entangled pair identifier (eprId).
3. Distributes each qubit of a pair and its eprId to the appropriate quantum network nodes.

A well-defined interface exists between the Quantum Network in Figure 8 and the logically distinct EPM function (Figure 9).

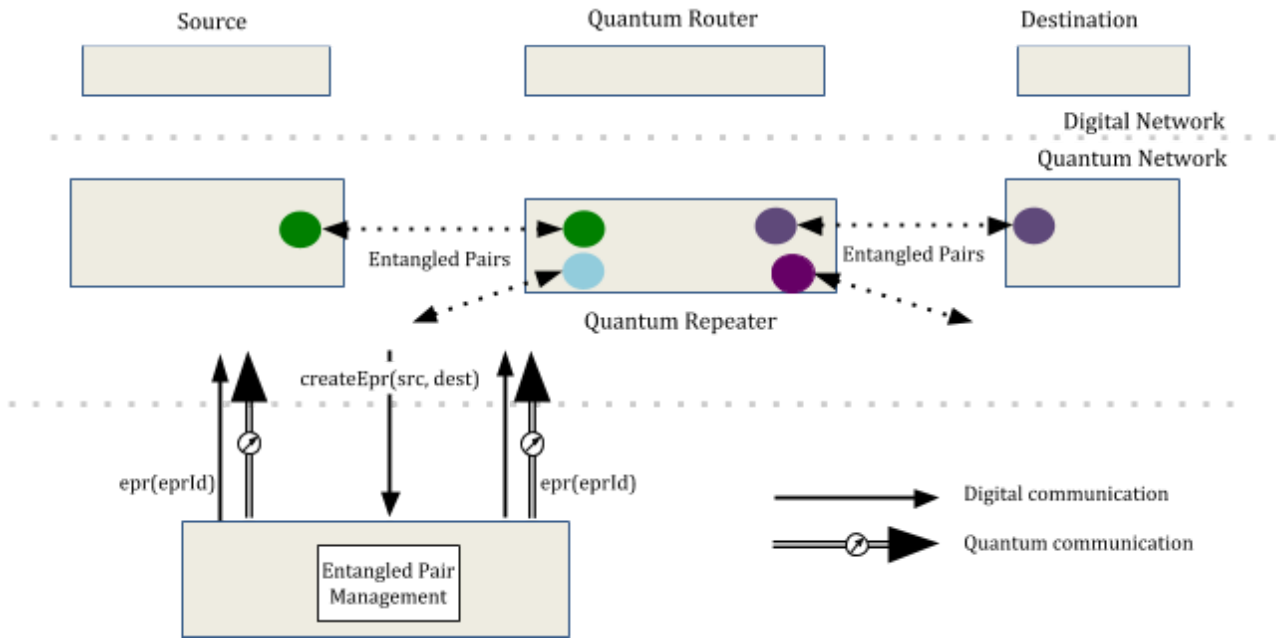

**Figure 9 - Entangled Pair Management**

1. *createEpr(src, dest)* causes EPM to create an entangled pair between the *src* and *dest* Quantum Network nodes. A variety of strategies for issuing *createEpr* are possible, e.g. when a pair is needed or in order to maintain a reserve of pairs between nodes.

2. EPM then:
    a. Creates an entangled pair.
    b. Sends one photon of the pair over a CW optical link to each of the requested Quantum Network nodes.
    c. Sends the entangled pair label (*eprId*) over a digital interface to the Quantum Network nodes.

## Entangled Web Prototype

We have created a prototype version of the Quantum Router to investigate the functions of the digital and quantum forwarding planes and the interface between them. The quantum network protocol was verified using a JavaScript quantum computer simulator [JSQUBITS]. The digital network functions and the interface to the quantum network were written in JavaScript and node.js [NODEJS]. Below is the simulation output for the example in Figure 8. A next step will be to replace the quantum simulator with physical teleportation and repeater nodes.

```
Source Node
    get𝑞𝑢𝑏𝑖𝑡()
        return (0.4091)|0> + (0.9125)|1>
    teleport(qubit: (0.4091)|0> + (0.9125)|1>, nextHop: QIR)
        Entangled Pair ID: 0, state: (0.7071)|00> + (0.7071)|11>
        Bell State: (0.2893)|000> + (0.2893)|011> + (0.6452)|100> + (0.6452)|111>)
        return(epId: 0, bsmResult: 0)
    forward({"src":"Source","dest":"Destination","teleportResult":{"epId":0,"bsmResult":0}})
QIR
    qsr({"epId":0,"bsmResult":0})
        return(qubit: (0.4091)|0> + (0.9125)|1>)
    teleport(qubit: (0.4091)|0> + (0.9125)|1>, nextHop: Destination)
        Entangled Pair ID: 1 state: (0.7071)|00> + (0.7071)|11>
        Bell State: (0.2893)|000> + (0.2893)|011> + (0.6452)|100> + (0.6452)|111>)
        return(epId: 1, bsmResult: 3)
    forward({"src":"Source","dest":"Destination","teleportResult":{"epId":1,"bsmResult":3}})
Destination Node
    qsr({"epId":1,"bsmResult":3})
        return(qubit: (0.4091)|0> + (0.9125)|1>)
```

# SUMMARY AND NEXT STEPS

We have described how the addition of a novel quantum router to existing CW optics, quantum teleportation and quantum repeaters, can be used to transfer quantum state from any source to any destination over, in principle, unlimited distances. We have also made the case for the existence of an automated forwarding mechanism in order to realize the vision of a quantum Internet.

We then described the design of a quantum router that extends existing digital network forwarding concepts to the routing of quantum states. Our design defines a simple interface between the Digital and Quantum Networks and should be applicable to any quantum teleportation and repeater implementation.

While this quantum router enables the transmission of qubits from any source to any destination, its practicality depends on a number of steps that should become routine in such a network. Among them is the creation and distribution of entangled photons at will and high frequencies, effective mechanisms for qubit storage, and easy implementations of quantum teleportation. Given the pace of innovation in this field, it is not unreasonable to expect that a quantum Internet will eventually coexist with, and improve, the classical one we use today.